\documentclass[10pt,sigconf,letterpaper,nonacm]{acmart}

\usepackage[english]{babel}

\usepackage{blindtext}

\usepackage{placeins}

\usepackage{xcolor}

\usepackage{booktabs} 

\usepackage{varwidth}
\usepackage{enumitem}
\usepackage{float}
\usepackage{graphicx}
\usepackage{rotating}
\usepackage[font=footnotesize,format=plain,labelfont=bf,textfont=bf]{caption}
\usepackage[font=scriptsize]{subcaption}
\usepackage{amsfonts}
\usepackage{multirow}
\usepackage{listings}
\usepackage{balance}

\usepackage{mdwlist}
\usepackage{comment}
\usepackage[linesnumbered,ruled,vlined]{algorithm2e}
\usepackage{fancyvrb}
\SetKw{KwBy}{by}
\usepackage{relsize}
\usepackage{tabularx}
\newcolumntype{Y}{>{\centering\arraybackslash}X}
\usepackage{bold-extra}

\newcommand{\smartparagraph}[1]{\vspace{.02in}\noindent{\bf #1}}

\makecompactlist{itemize}{stditemize}
\usepackage{mdwlist}
\usepackage{multirow, makecell, colortbl}
\usepackage[most]{tcolorbox}

\tcbset{
    frame code={}
    center title,
    left=0pt,
    right=0pt,
    top=0pt,
    bottom=0pt,
    colback=gray!20,
    colframe=white,
    width=\dimexpr\columnwidth\relax,
    enlarge left by=0mm,
    boxsep=5pt,
    arc=0pt,outer arc=0pt,
    }
\usepackage{etoolbox}

\usepackage{cleveref}

\crefformat{section}{\S#2#1#3} 
\crefformat{subsection}{\S#2#1#3}
\crefformat{subsubsection}{\S#2#1#3}

\newcommand{\system}{$\texttt{ShortCut}$\xspace}

\newcommand{\new}{{P$4_{16}$}\xspace}

\newtheorem{observation}{Observation}

\begin{document}
\pagestyle{plain}
\title{Shortcutting Fast Failover Routes in the Data Plane}

\ccsdesc[500]{Computer systems organization~Reliability}

\author{Apoorv Shukla}
\affiliation{
	\institution{Huawei Munich Research Center}
	\country{Germany}
}
\email{apoorv.shukla1@huawei.com}

\author{Klaus-Tycho Foerster}
\affiliation{
	\institution{TU Dortmund}
	\country{Germany}
}
\email{klaus-tycho.foerster@tu-dortmund.de}

\begin{abstract}
In networks, availability is of paramount importance.
As link failures are disruptive, modern networks in turn provide Fast ReRoute (FRR) mechanisms to rapidly restore connectivity.
However, existing FRR approaches heavily impact performance until the slower convergence protocols kick in.
The fast failover routes commonly involve unnecessary loops and detours, disturbing other traffic while causing costly packet loss.
In this paper, we make a case for augmenting FRR mechanisms to avoid such inefficiencies. 
We introduce~\system that routes the packets in a loop free fashion, avoiding costly detours and decreasing link load.
\system achieves this by leveraging data plane programmability: when a loop is locally observed, it can be removed by \emph{short-cutting} the respective route parts.
As such, \system is topology-independent and agnostic to the type of FRR currently deployed.
Our first experimental simulations show that \system can outperform control plane convergence mechanisms; moreover avoiding loops and keeping packet loss minimal opposed to existing FRR mechanisms. 

\end{abstract}

\keywords{Fast Failover Routing, Reliability, P4, Data Plane Algorithms}

\maketitle

\setlength{\skip\footins}{1.2pc plus 5pt minus 8pt}
\begin{figure*}
\centering
\begin{subfigure}{0.33\linewidth}
  \centering
  \includegraphics[width=.99\linewidth]{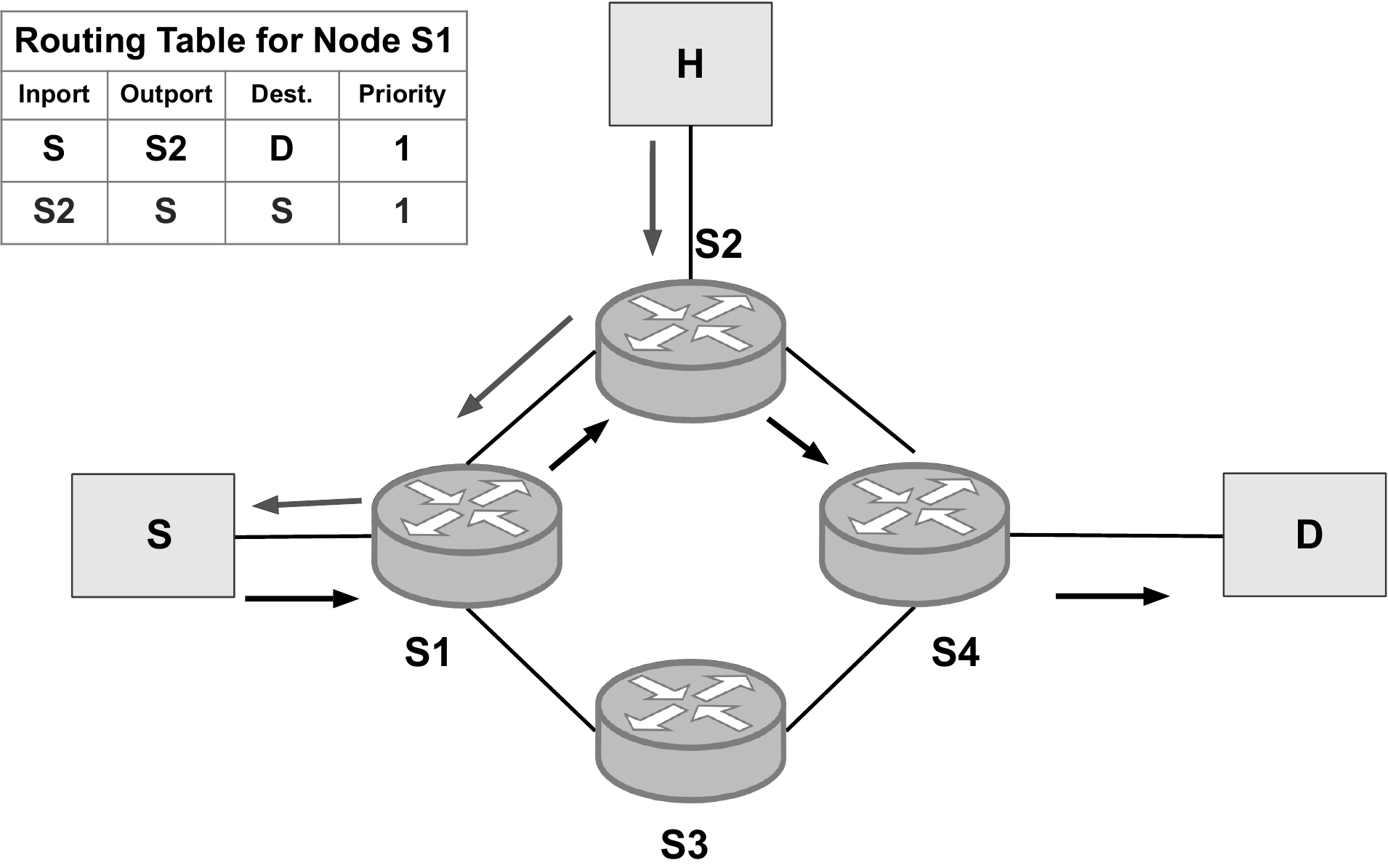}
  \subcaption{Without any failures}
  \label{fig:example-sub1}
\end{subfigure}
%
%
\begin{subfigure}{.33\linewidth}
  \centering
  \includegraphics[width=.99\linewidth]{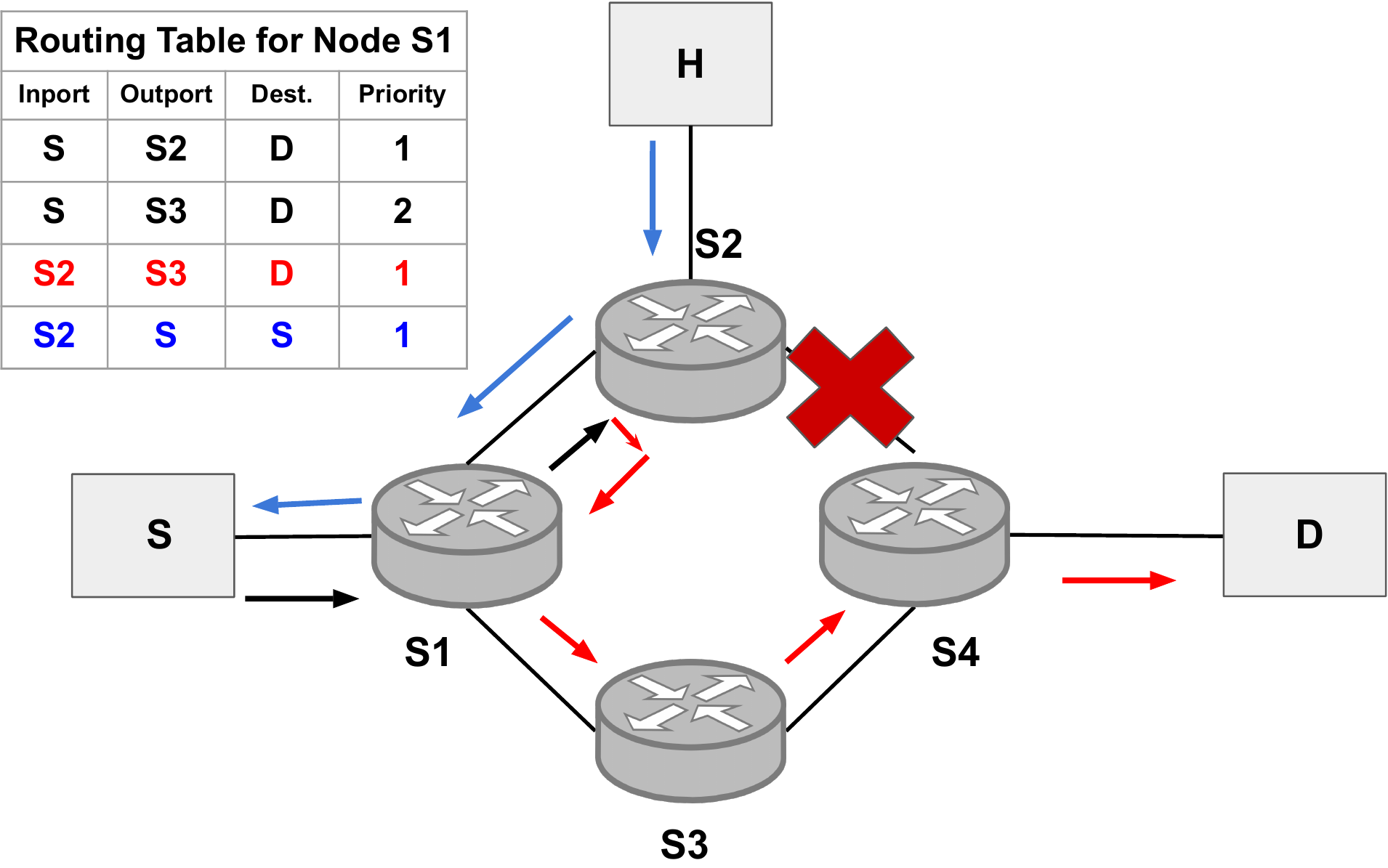}
  \subcaption{With FRR after a link failure}
  \label{fig:example-sub2}
\end{subfigure}
%
%
\begin{subfigure}{.33\linewidth}
  \centering
  \includegraphics[width=.99\linewidth]{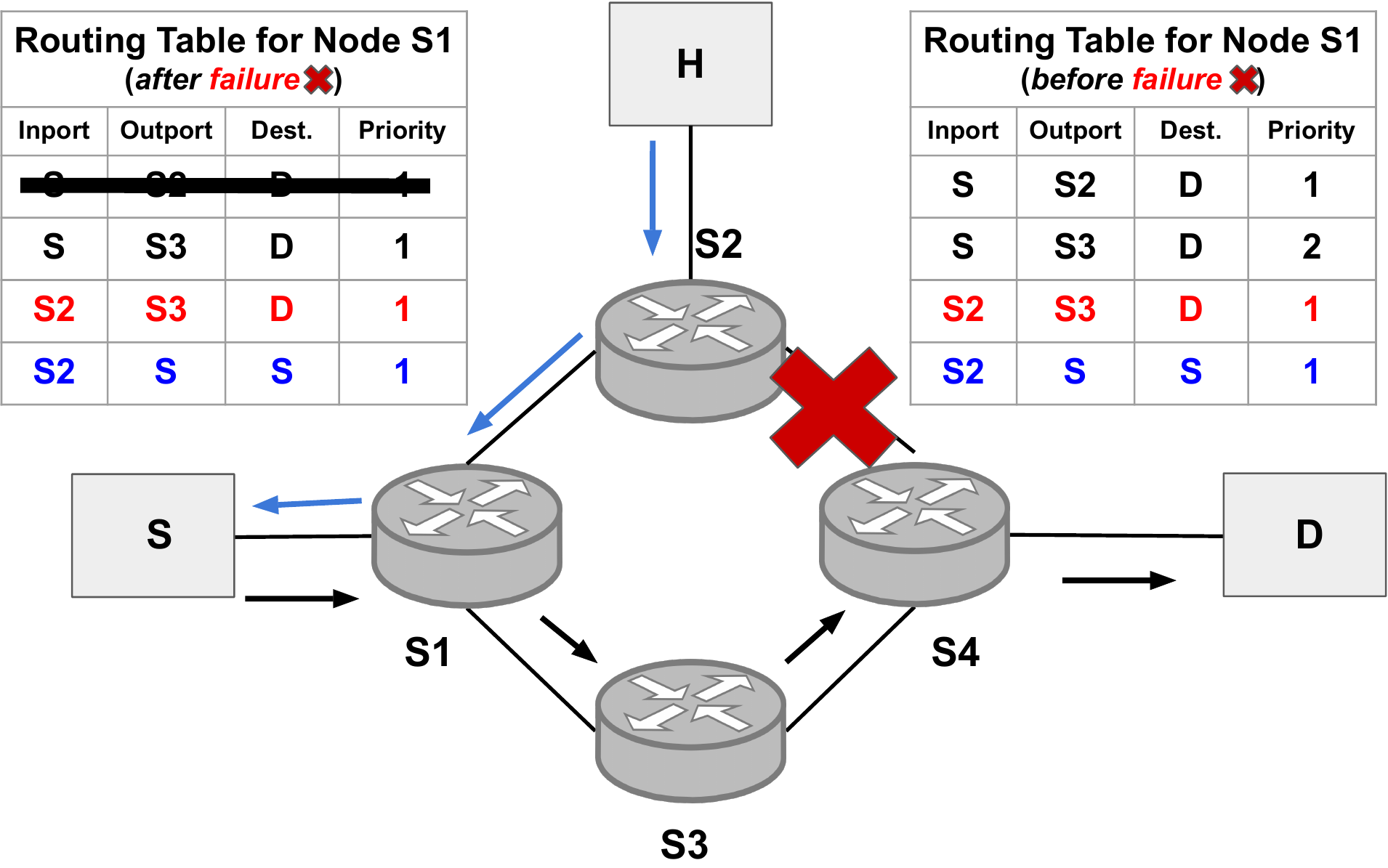}
  \subcaption{With \system\ after a link failure}
  \label{fig:example-sub3}
\end{subfigure}%
\vspace{-2mm}
\caption{Comparison of no FRR without failure, FRR, and \system\ }
\label{fig:example_new}
\vspace{2mm}
\end{figure*}

\section{Introduction}\label{sec:intro}

With the emergence of low-latency and high-bandwidth distributed applications~\cite{DBLP:conf/pam/ZilbermanGPBAW017} in datacenter or wide area networks, there is an ever-increasing pressure on the network operators in the form of stringent SLOs (Service Level Objectives) to ensure a peak performance in terms of availability, latency, bandwidth, and packet loss. However, unexpected failures (link/switch) are inevitable and happen regularly requiring a rapid action to ensure seamless connectivity without compromising on performance. 
A plethora of In-network Fast Reroute (FRR) approaches~\cite{DBLP:series/ccn/GanchevRCO20,frr-survey} have been developed entirely in the data plane to address such a problem. However, these approaches are slow, incur loops, trigger packet loss when routes become unavailable, to reroute traffic via a sub-optimal path~\cite{DBLP:conf/hotnets/LiuYSS11} which may be shared by other~traffic.

\smartparagraph{Control plane Convergence is slow}: In the light of failures, the \emph{global} control plane convergence is proven to be slow in seconds scale~\cite{DBLP:conf/nsdi/LiuPSGSS13} or even in some cases on a minutes scale~\cite{DBLP:conf/nsdi/HolterbachMADVV19}, adversely impacting the SLOs, and thus, business of network operators. The reason for slow control plane convergence is attributed to detecting failures, notifying switches of failures, recomputing new paths, and updating forwarding states depending on switch control plane design~\cite{francois2005achieving} accordingly. Therefore, in order to meet the SLOs, \emph{local} FRR mechanisms~\cite{DBLP:conf/conext/ChiesaSABKN019} have been deployed on the data plane for fast reaction to unexpected and crippling failures.

The conventional wisdom is to proactively install backup rules on the switches which take priority when the failure happens. Therefore, a hierarchical control plane design with a global control plane and local reactive control on the data plane has emerged as a popular~approach.

\smartparagraph{Navigating the FRR landscape of FRR:} 
Local FRR~\cite{DBLP:conf/podc/FeigenbaumGPSSS12,DBLP:journals/ton/ChiesaNMGMSS17,chiesa2016exploring,DBLP:conf/infocom/ElhouraniGR14,LFA,DBLP:conf/infocom/FoersterP0T19,DBLP:conf/infocom/YangLSLZ14,DBLP:conf/infocom/FoersterKP0T21,DBLP:conf/srds/FoersterKP0T19,DBLP:conf/dsn/FoersterKP0T19} can react almost immediately~\cite{DBLP:conf/conext/ChiesaSABKN019} to failures by proactively installing reroute rules, \emph{e.g.}, of lower priority.
As such they are the fastest failover schemes, attempting to always maintain connectivity, but can only do so with some downsides.
First, in many scenarios, it is impossible to protect against more than a single failure~\cite{DBLP:conf/podc/FeigenbaumGPSSS12,DBLP:journals/ton/ChiesaNMGMSS17,cs6529}.
Second, reminiscent of graph exploration~\cite{DBLP:journals/comcom/BorokhovichRSS18}, packets probe for working paths, introducing long detours.
%
%

%
Local FRR can also be achieved by means of packet header modification or encapsulation~\cite{I-Dietf-rtgwg-segment-routing-ti-lfa,rfc4090,DBLP:conf/sigcomm/LakshminarayananCRASS07}. Here however one has the drawbacks of needing custom-tailored protocols, potentially disturbing other network functions due to, \emph{e.g.}, header changes, issues due to increased packet sizes, \emph{e.g.}, with the MTU, and encountering reroute loops as well.
Moreover, there is FRR that leverages control plane convergence ideas in the data plane itself.
For example, \emph{DDC}~\cite{DBLP:conf/nsdi/LiuPSGSS13} utilizes link reversal algorithms~\cite{DBLP:journals/tcom/GafniB81,DBLP:series/synthesis/2011Welch} to provide connectivity, as long as the network is not partitioned.
\emph{Blink}~\cite{DBLP:conf/nsdi/HolterbachMADVV19}, on the other hand, tracks TCP disruptions to switch to backup routes.
While FRR implementation in the data plane, opposed to the control plane, significantly speeds up recovery, these mechanisms cannot provide the same rapid protection as local FRR schemes: \emph{e.g.}, \emph{Blink}~\cite{DBLP:conf/nsdi/HolterbachMADVV19} needs disruptions to occur first and link reversal algorithms, as in~\emph{DDC}~\cite{DBLP:conf/nsdi/LiuPSGSS13}, face \emph{quadratic convergence time} in the worst case~\cite{DBLP:conf/spaa/BuschST03}.
Still, the above convergence mechanisms also cover a wider range of multiple link failure scenarios, unobtainable by local FRR~\cite{DBLP:conf/podc/FeigenbaumGPSSS12,cs6529}.
%

%
%

\smartparagraph{Problem Statement.}
In this paper, we pose the following question: \emph{Local FRR maintains reachability by paying the price of additional delay and network load. Can we locally and rapidly remove those inefficient detours, before relatively slow convergence protocols kick~in?}

\smartparagraph{Solution Design.}
To this end we propose \system, which augments FRR by locally removing reroute loops while maintaining protection to a link failure, largely agnostic to underlying FRR.
\system is enabled by data plane methods, but unlike convergence methods, is purely local, without (implicit) message exchange by, \emph{e.g.}, link reversals, and furthermore does not rely on packet loss TCP signaling, maintaining immediate protection.
Hereby \system\ expands the design space of hierarchical FRR and convergence schemes, placing itself as an intermediate layer between both.

\noindent\textbf{\system leverages programmable data planes:} 
The P4 language~\cite{bosshart2014p4,p416specs} enables the programmability and customization of data plane functionalities in network devices. P4, an open-source language, allows programming of packet forwarding planes, and is increasingly supported by a panoply of network vendors. Via the P4 language, one can define in P4 programs the instructions for processing the packets, e.g., how the received packet should be read, manipulated, and forwarded by a network device, e.g.,~a~switch. 

Speaking of ``local'' fast reroute, P4 programs offer the required platform to fast reroute the packets on desired links at line rate when the link failure occurs while avoiding crippling loops and costly packet loss. Such ``local'' capability is crucial as the ``global'' control plane convergence mechanisms are slow~\cite{DBLP:conf/nsdi/LiuPSGSS13,DBLP:conf/nsdi/HolterbachMADVV19} to react to data plane events which require rapid action. Finally, when the global control plane mechanisms converge, they overwrite the local \system. We observe such hierarchical control as also, shown in~\cite{DBLP:conf/nsdi/LiuPSGSS13,DBLP:conf/sigcomm/YapMRPHBHKNJLRR17}, is crucial to meet the SLO targets. Our experimental simulation endorses our position as \system\ outperforms the global control plane convergence. Furthermore, we show that \system\ avoids costly loops and thereby load-induced packet loss~\cite{DBLP:journals/ton/BorokhovichPST18}, unlike existing FRR~\cite{DBLP:conf/infocom/ElhouraniGR14,chiesa2016exploring,DBLP:journals/ton/ChiesaNMGMSS17,DBLP:conf/infocom/FoersterP0T19,DBLP:conf/podc/FeigenbaumGPSSS12,DBLP:conf/infocom/YangLSLZ14}. 



\smartparagraph{ Contributions.} Our main contributions are:\\
\noindent $\bullet$ 
We identify an untapped opportunity in local FRR mechanisms to shorten failover routes and propose \system, a data plane method leveraging it. Our method is largely agnostic to the deployed local FRR mechanism, and also leaves data packets unchanged, respectively does not require packet state on the switches. (\S\ref{sec:challenges})
\\
\noindent $\bullet$ 
We prove correctness and efficiency of \system, \emph{i.e.}, single link failure protection under shorter (loop-free) routes. Moreover, we show that \system\ is realizable without additional communication, \emph{i.e.}, just by observing the data plane.~(\S\ref{sec:correctness})
\\
\noindent $\bullet$ 
We conduct an experimental evaluation of a \system\ prototype: \system strongly outperforms control plane convergence mechanisms, removing FRR loops and keeping packet loss minimal.~(\S\ref{sec:prototype})
\\
\noindent $\bullet$ 
We rigorously discuss FRR mechanisms and their interplay with \system, charting the landscape of FRR in depth.%
~(\S\ref{sec:discussion})

\section{Motivation and Background}\label{sec:challenges}
\noindent\textbf{The Control Plane is Slow.}
A cornerstone of FRR mechanisms is that reactions are immediate, ideally always maintaining logical connectivity. 
We cannot rely on instrumenting the control plane to this end, as ``\textit{the control plane typically operates at timescales several orders of magnitude slower than the data plane, which means that failure recovery will always be slow compared to data plane forwarding rates}''~\cite{DBLP:conf/nsdi/LiuPSGSS13}.
%


\subsection{Local Fast Reroute Mechanisms}
%
%
Hence, to react without delay to link failures, switches and routers must have the new routing already pre-computed, \emph{i.e.}, a mapping of incident faults to forwarding rules.
We give an example in Fig.~\ref{fig:example-sub1}, where the task is to route packets from a source $S$ to sink $D$, \emph{e.g.}, via the {\color{black} black} path S-S1-S2-S4-D.
When the link between S2 and S4 fails, a global view would change the routing at node S1, s.t.\ the new path is S-S1-S3-S4-D.
However, for immediate reactions, we cannot rely on the control plane, and hence only Node S2 (and S4) can change their behaviour immediately.
Here, the sole meaningful option at Node S2 is to bounce the packet back to its only neighbor S1, hoping that the packet reaches $D$.
%

%
At this point, the careful preprocessing of the network's topology by means of FRR comes into play.
State-of-the-art FRR leverages that nodes can send the same packet to different outports, based on the inport~\cite{DBLP:conf/infocom/YangLSLZ14}.
In other words, when the packet returns from Node S2 to Node S1, the Node S1 can now forward the packet to Node S3, from there to S4 and then to the destination.
Various methods have been proposed in this setting, such as (backtracking in~\cite{DBLP:conf/podc/FeigenbaumGPSSS12}) DAGs~\cite{DBLP:conf/hotnets/LiuYSS11}, partial structured networks~\cite{DBLP:conf/infocom/YangLSLZ14}, or arc-disjoint paths, trees, and arborescences~\cite{chiesa2016exploring,DBLP:journals/ton/ChiesaNMGMSS17}, all utilizing inport-awareness.

As such, FRR has maintained connectivity, by routing the packets along the black-{\color{red}red} path S-S1-S2-S1-S3-S4-D.
Notwithstanding, until the (slow) global convergence kicks in, this routing is inefficient, due to each packet looping once between S1 and S2.
What's worse, the rerouted packets will compete with the {\color{blue} blue} HS-flow, where we will lose up to $\approx 50\%$ of the total throughput.
Purely local and static FRR cannot overcome this inefficiency, the incident link fault state remains unchanged at Node S1; there are no incident~failures.

\subsection{Leveraging the Data Plane}
\vspace{-1mm}
We are motivated by the above scenario and hence aim at preserving $1)$ FRR connectivity guarantees while also $2)$ removing the inherent inefficiency of detour loops.
Our idea is to instrument the data plane to \emph{shortcut} unnecessary loops in FRR mechanisms, optimizing network performance until the control plane kicks in.
In more detail, we propose to observe the data plane, implicitly waiting for packets to traverse the same node twice, and then to remove routing~loops.
A straightforward approach to packet loop detection would be to remember packets or to mark them, which however comes with undesirable overhead in local storage or header expansion, the latter disturbing other network functions.
Rather, we propose to detect loops FRR implicitly, by means of different ports.
In Fig.~\ref{fig:example-sub1}, Node S1 expects packets, destined for D, to always arrive via S and to exit via Node S2, \emph{i.e.}, such packets arriving by Node S2 and leaving towards Node S3 implicitly signal a failure~downstream, as we explain next:
Node S1 deduces thereby that the returning packets traverse an unnecessary loop\footnote{they visit both S2 and S3 from S1, one after another} and \emph{shortcuts} the route by matching packets with inport S and destination D to the outport to Node S3.
From now on, no more packets will enter the loop S1-S2-S1, improving these links' utilization and in particular the latency of the flow's packets due to a shorter route. 
In this example, the shortcutted route is even already the route the control plane will converge to after some~time.
In more detail, a first packet from the SD-flow on the outport to Node S3 triggers Node S1 to change its routing, as shown in Fig.~\ref{fig:example-sub2}: at the inport from S, the top priority rule (to S2) is removed and the second priority rule (to S3) becomes the default.

\section{\system\ Details}\label{sec:correctness} 
%
%

%
We model the network as a graph $G=(V,E)$ with $n$ nodes (routers, switches, hosts) $V$ and $m$ directed links $E$.
We first define the routing for the failure-free case, \emph{i.e.}, without FRR.
Each packet in a flow $f$ is routed from a source $s=s(f) \in V $ to a destination $t = t(f) \in V $ along a simple path, \emph{i.e.}, a sequence of nodes without repetition.
We assume that the forwarding at a node $v$ (to an outport $v^o$ of $v$) is deterministic and may only match on $1)$ the flow's source%
\footnote{Our mechanism also works for destination-based routing via trees.}
and destination and $2)$ the incoming packet's port (inport $v^i$ ) at $v$.%
\footnote{\system also works for routing and FRR without usage of the inport, as we can then simply assume the forwarding is identical for all node inports.}
We next specify the FRR model and its routing for a link failure between any node pair $u,w$.
We will later discuss how to extend \system to further failure models.
Now, the forwarding at $u,w$ may also match on a third item, namely that the link $(u,w)$ is down.
Note that all other nodes are not aware of this failure and hence leave their routing unchanged.
Now, FRR may route the packets along a walk, \emph{i.e.}, node repetitions are allowed, but due to the deterministic forwarding behaviour, link repetitions lead to the packet not reaching the destination---as we assume packets are not modified by FRR.
We assume that each node $v$ has an ordered priority of outports $v^o_1, \ldots, v^o_k$ for a given flow (or destination), where $v^o_1$ is the default outport without failures.
Herein, the forwarding from each inport $v^i$ may implement a part of this priority list, \emph{i.e.}, $v^o_j, \ldots, v^o_k$, for $1 \leq j  \leq k$.%
We give an \textbf{example} for FRR with link-disjoint spanning arborescences\footnote{An arborescence is a directed tree oriented towards its root.}~\cite{DBLP:conf/infocom/ElhouraniGR14}.
In this FRR scheme, the idea is to first try to route on the first arborescence (tree), if a failure is encountered, then to switch to the second one, and so on.
To this end, at each node $v$, the outport $v^o_1$ corresponds to the first arborescence, $v^o_2$ to the second, and so on, where the inport of the first arborescence starts the priority list with $j=1$, the inport of the second arborescence with $j=2$ etc.
\vspace{-1mm}


\subsection{\system\ Mechanism}\label{subsec:flow-frr-mechanism}
%
At each node $v \in V$, \system\ performs the following operation continuously: for each inport $v^i$ and flow $f$, if packets from $f$ are sent through $1)$ an outport $v^o_h$ in the priority list $v^o_j, \ldots, v^o_k$ of inport $v^i$ for $f$ and $2)$ $v^o_h$ is not $v^o_j$, \emph{i.e.}, not its top priority, then $v^i$ removes the first items from its outport priority list until only $v^o_h, \ldots, v^o_k$ remains.
In other words, by observing that a lower priority outport is taken, corresponding inports make this outport their highest priority choice and remove all outdated higher priority outports in the process from their list.
Note that if an outport is not available due to link failure, it is considered as removed as~well.
%
%
\begin{observation}\label{obs:local}
\system operates locally at each node, without control plane messages or exchange between the nodes.
\end{observation}

\subsection{\system\ Properties}\label{subsec:flow-frr-guarantees}
 \system $1)$ maintains the underlying FRR reachability, $2)$ that the packet route turns into a loop-free sub-path of FRR, and $3)$ triggers within one end-to-end delay.

\begin{theorem}\label{thm:1-reach}
When the FRR scheme
maintains reachability under one link failure, then  \system maintains reachability  and changes the route to a loop-free FRR sub-path. 
\end{theorem}

\begin{proof}
\begin{figure}[t]
\centering
\begin{subfigure}{0.45\linewidth}
  \centering
  \includegraphics[width=.9\linewidth]{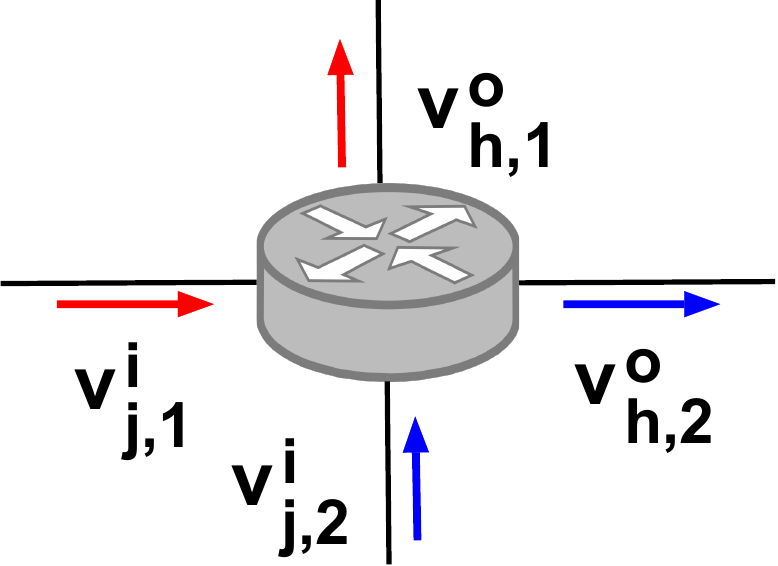}
  \subcaption{FRR rules without \system}
  \label{fig:example-proof1}
\end{subfigure}%
\begin{subfigure}{.45\linewidth}
  \centering
  \includegraphics[width=.9\linewidth]{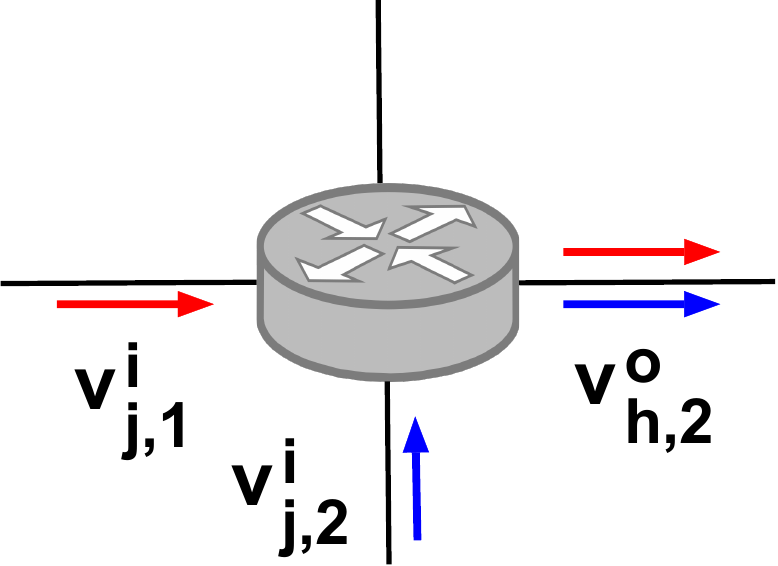}
  \subcaption{FRR rules with \system}
  \label{fig:example-proof2}
\end{subfigure}%
\vspace{-2mm}
\caption{Application of \system\ after a link failure}
\label{fig:example-proof}
\end{figure}
First, note that due to model assumptions, \system\ will not trigger before a link failure, as the standard outport has the highest priority on all nodes.
Second, we briefly investigate FRR after a link failure without \system, before the control plane convergence sets in: from the moment on the first packet hits the failed link (and rerouting is triggered), all following flow packets take the same route as the first flow packet.
This route may visit nodes multiple times, but it may visit each link only once due to the deterministic routing behaviour, else the packet would not reach the destination.
We now consider the impact of \system\ on FRR after a link failure at a node $v \neq t$ by case distinction over \#visits.
For one visit to $v$ per packet via $v^i_{j,1}$ to $v^o_{h,1}$, \system\ makes no change to the effective routing after FRR, even if some other inports also change their outport to $v^o_{j,1}$.
For two visits, first $v^i_{j,1}$ to $v^o_{h,1}$ and second $v^i_{j,2}$ to $v^o_{h,2}$, $v^o_{h,1}$ has a higher priority than $v^o_{h,2}$, i.e., $v^i_{j,1}$ changes its outport to $v^o_{h,1}$, but $v^i_{j,2}$ remains at $v^o_{h,2}$. 
Hence the loop beginning with $v^o_{h,1}$ and ending with $v^i_{j,2}$ is removed and from now on $v$ will only be visited once.
We refer to Fig.~\ref{fig:example-proof} for a visualization: in \ref{fig:example-proof1} the packet first visits and leaves the node $v$ via the red arrows, then later returns and leaves again via the blue arrows, forming a removable detour.
By application of \system\ in \ref{fig:example-proof2}, the red outgoing rule is adapted to the (later and blue) $v^o_{h,2}$, removing the detour.
As all packets exiting via $v^o_{h,2}$ reach the destination by assumption before application of \system, reachability is maintained, as we only remove the subpath between $v^i_{j,1}$ and $v^o_{h,2}$.

For more than two visits of $v$, say $\ell$, the argumentation is analogous, where eventually all relevant inports forward to the last outport~$v^o_{h,\ell}$, and visit $v$ only once.
Hence \system\ removes all loops while only using links also used by the FRR scheme.
Lastly, for a brief amount of time, there can be packets in flight that never hit the failure, but who might be impacted by \system.
Observe here that if they still use the standard routing on their remaining path, reachability is maintained, and else, they are detoured to (shortcutted) FRR rules, still reaching the destination.
\end{proof}
Observe that for the above routing change actions 
are all triggered as soon as a packet traverses the whole FRR path.
%
\begin{corollary}
After a link failure, all \system\ route change actions are triggered within one end-to-end delay.
\end{corollary}


\begin{figure}[t]
    \includegraphics[width=0.85\columnwidth]{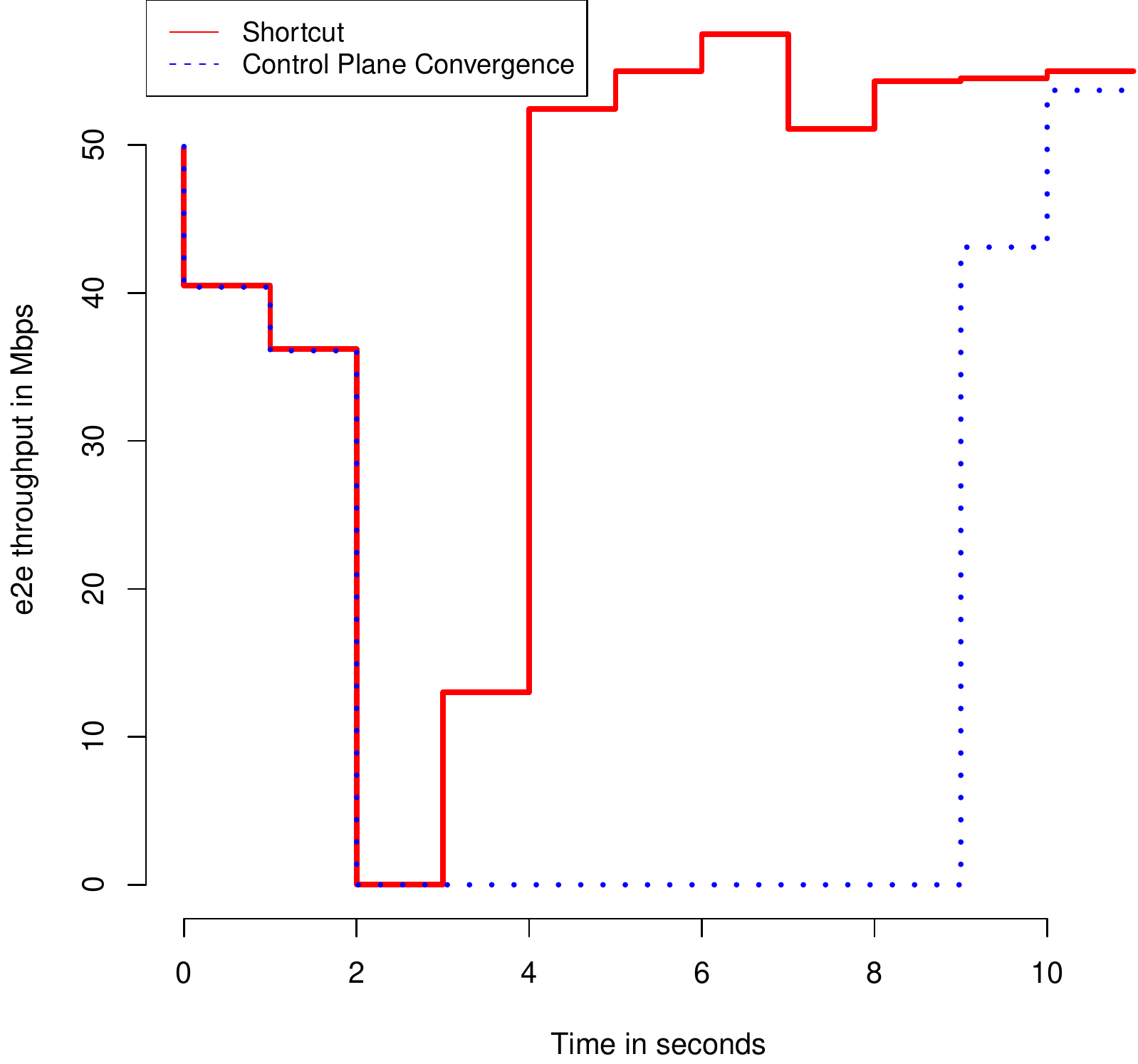}
    \vspace{-4mm}
    \caption{
    \system vs Control plane convergence w.r.t.\ mean end-to-end throughput over multiple TCP experiments. The link failure \textit{finally} comes into effect at 2 seconds, it takes seconds for the control plane convergence mechanisms to kick in, while \system\ routes traffic loop-free, maintaining tolerable packet loss.
    }
    \label{fig:exp}
\end{figure}

\section{Proof-of-Concept Evaluation}\label{sec:prototype}

In the following section, we demonstrate the effectiveness of a \system\ in P4 over control plane convergence mechanisms (routing protocols) in terms of throughput leveraging our P4 evaluation. In particular, we note that \system\ avoids loops and packet loss to ensure better throughput compared to control plane convergence mechanisms. 

For our experimental evaluation,
we choose the topology illustrated in Figure~\ref{fig:example_new} with preinstalled rules. We deploy an LPM (Longest Prefix match) P4 program~\cite{p4tut} written in \new~\cite{p416specs} in a Mininet~\cite{p4l} environment and leveraged iperf~\cite{iperf} for generating the end-to-end TCP traffic (between end hosts). Our evaluation leverages a centralized controller for simplicity.
Furthermore, P4 ensures fast routing on the network dataplane until the control plane convergence kicks in. Therefore, we leverage hierarchical control, \emph{i.e.}, global control with control plane convergence and local \system\ for fast reactive response to immediate dataplane~failures. 

To simulate a 1-link failure scenario under TCP traffic (see Figure~\ref{fig:example_new}), we use our custom Python-based script to ``fail and reroute'' traffic from S-S1-S2-S4-D to S-S1-S3-S4-D in the case of control plane mechanism and \system (See Figure~\ref{fig:example-sub3}). For our evaluation, we derive motivation from~\cite{DBLP:conf/nsdi/LiuPSGSS13, DBLP:conf/nsdi/HolterbachMADVV19}, where they observe that the control plane involves CPU operations which are significantly
slower (seconds scale) than the operations in the data plane (microseconds scale). It is noteworthy that until the control
plane mechanism kicks in, the unavailability of routes leading to packet loss will already occur. However, in \system, the rerouting happens at microsecond scale while avoiding loops and keeping packet loss to a minimum until switch-over to the new route happens. 

Note, in existing FRR approaches, there will be loops as the path taken will be S-S1-S2-S1-S3-S4-D and packet loss since, the rerouted flow (in {\color{red} red} in Figure~\ref{fig:example-sub2}) competes for the link capacity with another flow (in {\color{blue} blue} in Figure~\ref{fig:example-sub2}). 

Figure~\ref{fig:exp} illustrates the mean end-to-end throughput comparison against the control plane convergence mechanism before and after link failure over multiple experiments. We observe that \system\ outperforms the control plane convergence mechanisms with a rapid local reroute, avoiding loops and keeping packet losses minimal. We observe that the iperf tool reports that there is latency in the link failure. We note that link failure with zero throughput fully comes into effect at 2 seconds when we introduce the link failure at the zeroth second. Note, that \system's reroute is for a transient amount of time and is finally overwritten by the global control plane convergence.

\section{Discussion}\label{sec:discussion}
Beyond, \emph{e.g.}, arborescence-based FRR, we now provide a rigorous discussion on the applicability of \system\ to further mechanisms and scenarios, charting the local FRR landscape.
%

%

\smartparagraph{Greedy FRR.}
Some FRR mechanisms use a \emph{greedy} approach to circumvent link failures.
For example, in some regular graph topologies such as the 2-dimensional torus~\cite{DBLP:journals/ccr/FoersterPST18}, it suffices to first try to go closer to the destination, and else ``shift'' slightly to other directions, then going closer again.
However, these mechanisms, due to their greedy nature, usually need to exclude the incoming edge as the next outport, as else the reroute can easily get stuck in a permanent loop; the exception is if the inport is the only remaining choice to reach the destination.
As such, there is no longer an ordered priority of outports at each node, as inports put themselves at the bottom of their priority
General examples of greedy FRR include directed acyclic graphs (DAGs)~\cite{DBLP:conf/podc/FeigenbaumGPSSS12} or partial structural networks (PSNs)~\cite{DBLP:conf/infocom/YangLSLZ14}, where the quality of greedy FRR depends on the careful precomputation of the DAGs/PSNs. 
We can nonetheless let \system\ also augment greedy FRR.
If the (local) route is $v_1$-$v_2$-$v_1$, then, at $v_2$, bouncing back to $v_1$ was the best (greedy) choice.
Hence, if the packet were to return to $v_2$ later and would choose a different outport than to $v_1$, then this choice would already have been made at the earlier visit, due to the nature of greedy FRR, and as thus \system\ can set $v_1$ as the highest priority outport.
%
%

\smartparagraph{Circular FRR.}
Some FRR does not have a descending priority on alternate failover rules, but rather implements a circular scheme.
For example, Chiesa et al.~\cite{DBLP:conf/infocom/ChiesaNMPGMSS16} take on the concept of arborescence-based FRR in this way:
instead of just switching to the next arborescence upon hitting a failure, until the $k$-th (last) one is reached, they propose to switch in a circular fashion, \emph{i.e.}, from the $k$-th one back to the first one, and so on, instead of dropping the packet after packets on the $k$-th arborescence hit a link failure. 
Such FRR is problematic for \system\ as it is no longer clear from a local port view which outport corresponds to the ``last'' node visit of a~packet. 
However, in case circular FRR stabilizes to just a subset of outports, we can adapt \system.
We propose to add a small timer, corresponding to the FRR stabilization time, after which \system\ re-evaluates the local traffic; the purpose of the timer is to drain the network of flow packets emitted before the failure appeared. 
Now, \system\ keeps track of the first time a flow packet arrived by a non-standard inport $v^i$, and adapts the circular ordering to be non-circular where the outport chosen by $v^i$ is now the top of the list.
In other words, we rotate the circular ordering and remove its circular property.
An FRR scheme where this idea applies is failover routing on outerplanar graphs, where it suffices to fix an embedding and set all routing to be, \emph{e.g.}, clockwise~\cite{cs6529}.

\smartparagraph{Virtual Network Partition FRR.}
A further idea employed by some FRR mechanisms is to virtually partition the network $G=(V,E)$ into link-disjoint parts $G_1,\ldots,G_k$.
Then, in each part $G_i$, the FRR scheme attempts to reach the destination, continuing with $G_{i+1}$ if unsuccessful, and so on.
An example is to compute $k$ edge-disjoint source-destination paths, where the standard route is set in $G_1$~\cite{DBLP:conf/infocom/FoersterP0T19}.
\system\ can adapt to this setting by only updating the routing rules for each partition, \emph{e.g.}, if the inport of the standard route switches to a lower priority, only rules for $G_1$ are updated.
Hereby we remove routing loops from each partition. 
If the network partitions are strictly ordered as in, \emph{e.g.},~\cite{DBLP:conf/infocom/FoersterP0T19}, \system\ can jump locally to the next partition, removing even loops that span across network partitions.

\smartparagraph{Randomization, Duplication, and State.}
\system\ is deterministic and hence does not apply to FRR schemes utilizing, \emph{e.g.}, random walks~\cite{DBLP:journals/comcom/BorokhovichRSS18} or random arborescence switching~\cite{DBLP:conf/icalp/ChiesaGMMNSS16}: there is no longer a mapping which outport is ``last''.
Similarly, \system\ is also not designed for FRR with packet duplication~\cite{DBLP:journals/ton/ChiesaNMGMSS17}, as these schemes implicitly accept that some duplicated packets will not reach the destination, and hence locally we cannot distinguish which ones were successful.

Moreover, \system\ does not keep state in the packets or at the nodes and hence is not directly compatible with such FRR, \emph{e.g.}, the rotor-router mechanism~\cite{DBLP:journals/algorithmica/BampasGHIKKR17}, which essentially moves the memory of an exploring agent to the nodes.
However, if the state is strictly ordered, then \system\ can be extended in this direction, \emph{e.g.}, when using $\log k$ bits for counting changes between $k$ arborescence~\cite{DBLP:conf/infocom/ElhouraniGR14}.

\smartparagraph{Multiple Failure FRR.}
\system\ is designed with a single link failure in mind, as we expect the control plane to deploy new rules until a new failure appears.
However, extending \system\ to multiple link failures is difficult, as \system\ is largely agnostic to the underlying FRR.
When \system\ removes the loops introduced by the FRR after the first link failure, it could very well be that exactly those loops guarantee destination reachability when further failures appear.
Notwithstanding, \system\ can incorporate the scenario of a whole node failing, taking all its attached links down with it.
The reason is that for a node failure, the FRR does not change over time, and \system\ removing FRR loops hence does not impact reachability of the flow's~destination.

\smartparagraph{Segment Routing and MPLS FRR.}
Conceptually, \system\ can also augment Segment Routing (SR) and MPLS FRR~\cite{rfc4090} schemes, treating the packet encapsulation or top label identically to flow routing rules.
However, \system\ will only remove routing loops for each individual encapsulated header respectively label, not across them.
It could be interesting to extend \system\ across the whole label stack.
For example, when using TI-LFA~\cite{I-Dietf-rtgwg-segment-routing-ti-lfa} for link protection, the segments to route around the failed link can intersect with the original route, and here \system\ could remove those loops across header segments/encapsulations, handling failure carrying packet FRR schemes~\cite{DBLP:conf/infocom/FoersterPCS18,DBLP:conf/sigcomm/LakshminarayananCRASS07} analogously.

\smartparagraph{Non-local \system\ implementation.}
\system\ only requires observation of the data plane at each node individually, without communication between the nodes and/or a logically centralized controller, recall Observation~\ref{obs:local}.
Still, \system\ requires some ability to actually change the routing at a node upon being triggered by the data plane.
While we believe future, \emph{e.g.}, P4 extensions or custom programs could be used to this end, a direct solution would be to proceed analogously as proposed by Ngyuen et al.~\cite{DBLP:conf/sosr/NguyenCC17}: each node has its own (low-cost) controller, allowing to implement routing table updates near instantaneously.
Alternatively, a distributed control plane with, \emph{e.g.}, multiple controllers could be leveraged~\cite{DBLP:conf/sigcomm/SchmidS13}, or even a classic centralized controller setup: while a new routing configuration is prepared, the controller could rapidly issue the simple updates needed by \system.

\smartparagraph{Temporary Failures.}
Some link failures are only temporary, \emph{e.g.}, link-flaps due to protocol issues~\cite{DBLP:conf/cloud/PotharajuJ13} or optical reconfiguration~\cite{DBLP:conf/sigcomm/SinghGFFG18}.
In these cases \system does not automatically switch back to the now again available route, and it would be interesting to study trade-offs involving delays and probing~\cite{DBLP:conf/nsdi/HolterbachMADVV19}, before the control plane takes over.
It it would moreover be interesting to investigate the interplay of \system with temporary inconsistencies due to bugs or outdated control plane views~\cite{DBLP:conf/infocom/ShuklaHVHSH0F21,DBLP:journals/jsac/ShuklaFZHS20,DBLP:journals/tnsm/ShuklaSSCZF20}, respectively during network updates~\cite{DBLP:journals/comsur/FoersterSV19}, \emph{e.g.}, separated into fast-paced rounds~\cite{DBLP:conf/sigcomm/ShuklaSFLDS16}, and with route verification in P4~\cite{p42021}.

\section{Related Work}\label{sec:rel}
Resilient routing has been widely studied~\cite{DBLP:series/ccn/GanchevRCO20}, especially for fast recovery and reroute mechanisms~\cite{frr-survey}.
We next focus on $1)$ local fast reroute (FRR) mechanisms, which covers statically pre-installed failover rules, and $2)$ non-local recovery mechanisms by means of (control/data plane) convergence. 

\smartparagraph{Static local FRR mechanisms} have the advantage that routing is deterministic, that no additional (packet) memory is required on the nodes (or alternatively, tagging of the packet), and in particular no message exchange is required.
Chiesa et al.~\cite{chiesa2016exploring,DBLP:journals/ton/ChiesaNMGMSS17} use link-disjoint destination-rooted spanning arborescences to this end, where the resilience is related to the number of arborescences the network supports; after a failure the next arborescence is chosen, see also~\cite{DBLP:conf/infocom/StephensC16}.
\emph{CASA}~\cite{DBLP:conf/infocom/FoersterP0T19} investigates here how to minimize the load under rerouting in arborescences, also looking into edge-disjoint paths---extended in~\cite{treeancs}.
Conceptually, \emph{CASA} takes some inspiration from U-Turn~\cite{atlas2006u}, which extends LFA protocols to multi-hop repair paths, by pushing the packet back to a point where it can potentially reach the destination, possible iteratively~\cite{DBLP:conf/iwqos/ZhangWB13}. 
However these and the next works provide no mechanisms on how to short-cut packets traversing nodes more than once.
Various further works~\cite{DBLP:conf/podc/FeigenbaumGPSSS12,DBLP:conf/infocom/LeeYNZC04,DBLP:journals/ton/NelakuditiLYZC07,DBLP:conf/iwqos/NelakuditiLYZ03,wang2007ip,DBLP:conf/infocom/ZhongNYLWC05} consider how to protect against only single failures, but in contrast work without any topology assumptions.

Different from worst-case guarantees, Yang et al.~\cite{DBLP:conf/infocom/YangLSLZ14} propose a version of greedy FRR, where packets try to get closer to the destination, or at least not increase their distance. 
%
%

%
\system\ is complementary to the above FRR mechanisms and can augment them by locally removing routing loops induced by rerouting, turning the packet route into a path, and hence reducing packet delay and the congestion of other links in said loops.
We are not aware of other works that operate in this setting, \emph{i.e.}, deterministic without tagging, probing, state, or message exchange, as an intermediate between local FRR and convergence schemes.

Furthermore, local FRR that relies on randomization~\cite{DBLP:conf/icalp/ChiesaGMMNSS16,DBLP:journals/comcom/BorokhovichRSS18} or on state to remember packets~\cite{DBLP:journals/algorithmica/BampasGHIKKR17}.
However, both can be problematic in practice and are non-standard, requiring extra memory and randomization beyond hashing, while causing packet reordering.
Additionally, some local FRR moves the additional memory into the packets, \emph{i.e.}, by means of failure carrying packets~\cite{DBLP:conf/hotnets/StephensCR13,DBLP:conf/sigcomm/LakshminarayananCRASS07}, MPLS~\cite{rfc4090}, segment routing~\cite{I-Dietf-rtgwg-segment-routing-ti-lfa}, or as general rewritable header\footnote{There are also solutions that use header space to detect loops in the (P4) data plane, for example the recent work by Kucera et al.~\cite{DBLP:conf/conext/0004BKAYM20}.} space~\cite{DBLP:journals/ton/ChiesaNMGMSS17,DBLP:conf/infocom/ElhouraniGR14}.
\system can conceptually be expanded in this latter direction, under the assumption that the memory content induces a strict ordering and/or if the whole label stack can be analyzed. 
We note that, unless failures are (implicitly) added to the packets~\cite{DBLP:conf/sigcomm/LakshminarayananCRASS07}, it is unclear how to extend, \emph{e.g.}, segment routing~\cite{I-Dietf-rtgwg-segment-routing-ti-lfa} beyond protection for a single link in general~\cite{DBLP:conf/infocom/FoersterPCS18}.

\smartparagraph{Control plane convergence methods} to optimize routing however are well established, such as classical distance-vector and link-state algorithms, also centralized SDN methods~\cite{DBLP:series/ccn/GanchevRCO20}.
Nonetheless, they all suffer from significant delays in comparison to data plane speeds~\cite{DBLP:conf/nsdi/LiuPSGSS13}.\footnote{Reroute speed is significantly improved if each node has its own control plane~\cite{LFA}, only exchanging messages for convergence in a two-tier hierarchical approach. However, conceptually this can then be understood as (a precursor to) local FRR.}
As thus, the state of the art for best protection is a two-tier hierarchical approach, where local FRR provides immediate reroute, at the cost of non-optimal paths, followed by slower convergence protocols with a global view.
We refer to the recent book edited by Rak and Hutchison~\cite{DBLP:series/ccn/RH2020} for further discussions.

\smartparagraph{Data plane convergence methods} have recently taken off, allowing the possibility of similar recovery without invoking the control plane.
For example and notably, \emph{Blink} tracks TCP disruptions to quickly change paths after loss is indicated at the ingress.
Thus, it is notably faster (according to the authors, ``\textit{sub-second rerouting}''), but it also does not aim for immediate protection such as local FRR.
\emph{DDC}~\cite{DBLP:conf/nsdi/LiuPSGSS13} implements link reversal algorithms in the data plane~\cite{DBLP:journals/tcom/GafniB81,DBLP:series/synthesis/2011Welch}, and hence can recover from any non-partitioning failure set, faster than in the control plane, yet still cannot avoid slow (quadratic) recovery times inherent to link reversal~\cite{DBLP:conf/spaa/BuschST03}, and does not provide immediate protection as well. 
Ramadan et al.~\cite{DBLP:conf/podc/RamadanMDZ16} propose to speed up the convergence time by means of preorders and iteratively deactivating links.
Lastly, Chiesa et al.~\cite{DBLP:conf/conext/ChiesaSABKN019} proposed FRR primitives for programmable data planes which can complement~\system, and Stephens et al.~\cite{DBLP:conf/sosr/StephensCR16} investigate how to scale FRR rules by compression.

\section{Conclusion}\label{sec:conclusion}
%
We studied the fundamental question of how to improve fast failover routes in the data plane.
Current fast reroute mechanisms maintain reachability by means of detours, where extra delay and additional link load persists until the relatively slow control plane kicks~in.

Our system, \system, leverages the observation that local fast failover routes often contain transient loops, which can be shortcut in the data plane. 
By removing such unnecessary detours locally, \system can rapidly improve fast reroute quality while maintaining reachability and protection guarantees.
Herein, \system is not a replacement for already implemented failover protection, but rather augments them, being largely agnostic to the local fast reroute mechanisms in place.
As such, the protection guarantees of current (and future) fast reroute implementations are maintained, with \system improving the network performance until the control plane reconverges.

Our experimental simulations of a \system\ prototype showcases feasibility and benefits over slower control plane convergence mechanisms, removing FRR loops and their induced packet loss.
We moreover discuss the existing FRR landscape and their interplay with \system\ in-depth, charting future directions and extensions. 
Conceptually, \system\ expands the design space of hierarchical fast reroute and recovery mechanisms, placing itself as an intermediate between rapid FRR and slow convergence.

\vspace{1mm}
\noindent\textbf{Outlook.}
Even though \system\ is largely oblivious to the underlying local FRR mechanism, and hence has the advantage of being widely applicable, it could be interesting to provide more direct integration into FRR and recovery schemes, trading off generality versus performance.
As next steps, we plan to investigate the extension of \system to multiple failures, respectively to show where such extensions are infeasible, and to expand our proof-of-concept implementation, along with large-scale simulations.

\vspace{1mm}
\noindent\textbf{Acknowledgments and Bibliographical Note.}
This paper appears~\cite{shortcut} at the ACM/IEEE Symposium on Architectures for Networking and Communications Systems 2021 (ANCS'21) and we thank the anonymous reviewers for their valuable feedback. 

\vspace{1mm}
\noindent\textbf{Reproducibility.}
Our experimental evaluation will be made available at \url{https://github.com/Apoorv1986/Shortcut}.

\balance
\bibliographystyle{ACM-Reference-Format}
\bibliography{refs}
\renewcommand{\thesubsection}{\Alph{subsection}}

\end{document}